\documentclass[twocolumn,prb,aps]{revtex4}
\usepackage{amsmath,amssymb}
\usepackage{graphicx}

\graphicspath{{./Figures/}}

\newcommand{\eref}[1]{Eq.~(\ref{#1})}
\newcommand{\erefs}[2]{Eqs.~(\ref{#1})~and~(\ref{#2})}

\newcommand{\fref}[1]{Fig.~\ref{#1}}

\newcommand{\sref}[1]{Section~\ref{#1}}

\newcommand{\aref}[1]{Appendix~\ref{#1}}

\newcommand{\erf}{\ensuremath{\mathop{\rm erf}}}
\newcommand{\erfc}{\ensuremath{\mathop{\rm erfc}}}

\newcommand{\vect}[1]{\mathbf{#1}}

\newcommand{\vdiff}[2]{\left|\vect{#1} - \vect{#2}\right|}

\newcommand{\eg}{\emph{e.g.}}

\newcommand{\Vr}{\ensuremath{\mathcal{V}_{\rm R}}}

\newcommand{\lmf}{LMF}

\newcommand{\sig}{\ensuremath{\sigma}}

\newcommand{\vs}{\ensuremath{v_0(r)}}
\newcommand{\vl}{\ensuremath{v_1(r)}}

\newcommand{\spce}{SPC/E}

\newcommand{\kT}{\ensuremath{{\rm k_BT}}}
\newcommand{\allpos}{\vect{\overline{R}}}

\begin{document}


\title[Thermodynamics for Short-Ranged Truncations of Coulomb Interactions]{Accurate Thermodynamics for Short-Ranged Truncations of Coulomb Interactions in Site-Site Molecular Models}


\author{Jocelyn M. Rodgers}
\altaffiliation[Present address: ]{Materials Science Division, Lawrence Berkeley National Laboratory, Berkeley, CA 94720}
\affiliation{Institute for Physical Science and Technology, University of
  Maryland, College Park, Maryland 20742}
\affiliation{Chemical Physics Program, University of Maryland, College Park, Maryland 20742}

\author{John D. Weeks}
\email{jdw@ipst.umd.edu}
\affiliation{Institute for Physical Science and Technology, University of
  Maryland, College Park, Maryland 20742}
\affiliation{Department of Chemistry and Biochemistry, University of Maryland,
  College Park, Maryland 20742}


\date{\today}

\begin{abstract}
Coulomb interactions are present in a wide variety of all-atom
  force fields.  Spherical truncations of these
  interactions permit fast simulations but are problematic
  due to their incorrect thermodynamics. Herein we demonstrate
  that simple analytical corrections for the
  thermodynamics of uniform truncated systems are possible.  In
  particular results for the SPC/E water model treated with
  spherically-truncated Coulomb interactions suggested by local
  molecular field theory~[Proc. Nat. Acad. Sci. USA {\bf 105}, 19136
  (2008)] are presented. We extend results developed by
  Chandler~[J.~Chem.~Phys.\ {\bf 65}, 2925 (1976)] so that we may
  treat the thermodynamics of mixtures of flexible charged and
  uncharged molecules simulated with spherical truncations.
  We show that the energy and pressure of spherically-truncated
  bulk SPC/E water are easily corrected using exact second-moment-like
  conditions on long-ranged structure.  Furthermore, applying the
  pressure correction as an external pressure removes the density
  errors observed by other research groups in NPT simulations of
  spherically-truncated bulk species.
\end{abstract}


\maketitle

\section{Introduction \label{sxn:intro}}

Most classical intermolecular potential models assign effective point
charges to intramolecular interaction sites to describe charge
separation in polar molecules and the ability to form hydrogen
bonds~\cite{DuanWuChowdhury.2003.A-point-charge-force-field-for-molecular-mechanics,MacKerellBashfordBellott.1998.All-Atom-Empirical-Potential-for-Molecular-Modeling}.
 Thus, even in purely neutral systems, charge-charge interactions
remain important and expensive components of molecular simulations,
usually dealt with via Ewald summations or some other lattice-sum-like
technique~\cite{FrenkelSmit.2002.Understanding-Molecular-Simulation:-From-Algorithms}.

Recently there has been renewed interest in
spherically truncating the $1/r$ interaction and neglecting the
long-ranged components beyond a specified cutoff
radius~\cite{FennellGezelter.2006.Is-the-Ewald-summation-still-necessary-Pairwise,WolfKeblinskiPhillpot.1999.Exact-method-for-the-simulation-of-Coulombic-systems,Nezbeda.2005.Towards-a-Unified-View-of-Fluids,IzvekovSwansonVoth.2008.Coarse-graining-in-interaction-space:-A-systematic-approach}.
This permits fast and efficient simulations that scale
linearly with system size.
However, spherical truncation are problematic to implement for Coulomb
interactions. While many groups have found that accurate local pair
correlation functions in uniform systems may be obtained by a variety of spherical
truncations of
$1/r$~\cite{FennellGezelter.2006.Is-the-Ewald-summation-still-necessary-Pairwise,WolfKeblinskiPhillpot.1999.Exact-method-for-the-simulation-of-Coulombic-systems,Nezbeda.2005.Towards-a-Unified-View-of-Fluids,IzvekovSwansonVoth.2008.Coarse-graining-in-interaction-space:-A-systematic-approach,HummerSoumpasisNeumann.1994.Computer-simulation-of-aqueous-Na-Cl-Electrolytes},
two common and valid objections to spherical truncations remain:
\begin{enumerate}
\item they fail for structural and electrostatic properties in
  nonuniform systems, $e.g.$, systems with point charges confined between
  walls~\cite{FellerPastorRojnuckari.1996.Effect-of-Electrostatic-Force-Truncation-on-Interfacial,Spohr.1997.Effect-of-Electrostatic-Boundary-Conditions-and-System},
  and
\item in uniform systems, the thermodynamics predicted by such
  truncations~\cite{FrenkelSmit.2002.Understanding-Molecular-Simulation:-From-Algorithms}
  and even the bulk densities in NPT
  simulations~\cite{IzvekovSwansonVoth.2008.Coarse-graining-in-interaction-space:-A-systematic-approach,ZahnSchillingKast.2002.Enhancement-of-the-wolf-damped-Coulomb-potential:}
  are known to be inaccurate.
\end{enumerate}

Recently we overcame the first objection, showing that local molecular
field (\lmf)
theory~\cite{WeeksKatsovVollmayr.1998.Roles-of-repulsive-and-attractive-forces-in-determining,RodgersWeeks.2008.Local-molecular-field-theory-for-the-treatment}
provides an accurate path to structural properties in both ionic and
aqueous nonuniform systems using a spherical truncation of $1/r$ along
with a restructured external potential \Vr\ to account for the net
averaged effects of the long-ranged forces neglected in the spherical
truncation~\cite{ChenWeeks.2006.Local-molecular-field-theory-for-effective,RodgersKaurChen.2006.Attraction-between-like-charged-walls:-Short-ranged,RodgersWeeks.2008.Interplay-of-local-hydrogen-bonding-and-long-ranged-dipolar}.
Electrostatic properties are then also very accurately described~\cite{RodgersWeeks.2008.Interplay-of-local-hydrogen-bonding-and-long-ranged-dipolar,RodgersWeeks.2008.Water-in-strong-electric-fields:-Insights}.

\begin{figure}[tb]
  \centering
  \includegraphics[width=3.25in]{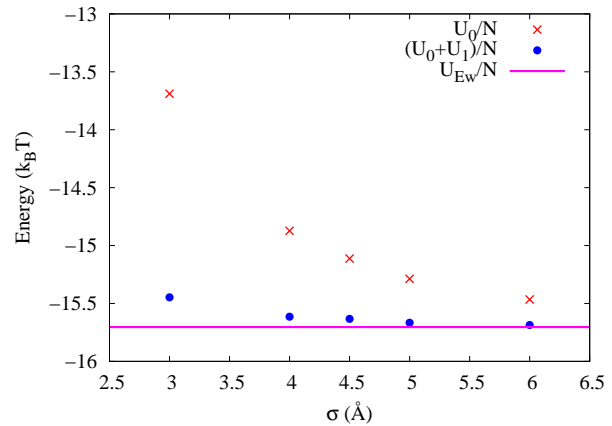}
  \caption{Plot of total potential energy without ($U_0/N$ in red
    crosses) and with the long-range correction ($U_0/N+U_1/N$ in blue
    circles) for the full range of $\sigma$ studied, representing
    greater inclusion of nearby core interactions.  The length
    $\sigma$ sets the scale for the smooth truncation of the Coulomb
    interactions as is explained further in \sref{sxn:lmf}.  Error bars are
    smaller than the data points. The Ewald determined energy is
    indicated by a horizontal line.}
  \label{fig:BulkECorr}
\end{figure}

Generalizing previous work for purely ionic
systems~\cite{ChenKaurWeeks.2004.Connecting-systems-with-short-and-long,DenesyukWeeks.2008.A-new-approach-for-efficient-simulation-of-Coulomb-interactions},
we show here that the \lmf\ framework also guides us to simple
analytic corrections for the energy and pressure of a general uniform
mixture of both polar and charged site-site molecules. As in thermodynamic
perturbation theory~\cite{HansenMcDonald.2006.Theory-of-Simple-Liquids}, we
can view a short-ranged truncation of $1/r$ as the reference system
for the fully interacting system. 
Our corrections are appropriate and accurate only for certain special well-chosen
reference systems as discussed below, which we refer to as ``mimic systems''. The
total energy and pressure of the full system is then given by the sums
\begin{align}
U_{\rm tot} &= U_0 + U_1 \nonumber \\ 
P_{\rm tot} &= P_0 + P_1.
\label{eqn:splitUP}
\end{align}
Typical simulations using such a spherical truncation of the $1/r$
interaction yield only $U_0$ and $P_0$, and our task is to determine the corrections
 $U_1$ and $P_1$ that would arise from an accurate treatment of the long-ranged interactions.

In particular, as shown in \fref{fig:BulkECorr} for bulk \spce\
water~\cite{BerendsenGrigeraStraatsma.1987.The-missing-term-in-effective-pair-potentials},
$U_0$ alone over a wide range of truncation distances parametrized by
the length $\sigma$ does not agree with the energy as calculated
using three-dimensional Ewald sums.  In contrast, any \sig\ of
3.0~\AA\ or greater reproduces the short-ranged pair correlations
predicted by Ewald sums quite
well~\cite{RodgersWeeks.2008.Interplay-of-local-hydrogen-bonding-and-long-ranged-dipolar,HuRodgersWeeks.2008.On-the-efficient-short-ranged-simulations-of-polar-molecular}.
As \sig\ increases, the energies are in better agreement with the
Ewald calculated values, but
noticeable differences remain even for large \sig.

Related problems arose in recent NPT ensemble simulations of
water~\cite{ZahnSchillingKast.2002.Enhancement-of-the-wolf-damped-Coulomb-potential:,IzvekovSwansonVoth.2008.Coarse-graining-in-interaction-space:-A-systematic-approach}.
Using Wolf
sums~\cite{WolfKeblinskiPhillpot.1999.Exact-method-for-the-simulation-of-Coulombic-systems},
researchers simulated systems quite similar to the truncated system
dictated by \lmf\ theory for $\sig \approx 5.0$~\AA\ (labeled
``DFS$_2$'' in
Ref.~\onlinecite{IzvekovSwansonVoth.2008.Coarse-graining-in-interaction-space:-A-systematic-approach})
and for a range of \sig\ in
Ref.~\onlinecite{ZahnSchillingKast.2002.Enhancement-of-the-wolf-damped-Coulomb-potential:}.
Each used the NPT ensemble with a pressure of 1~atm. They found
generally good structural agreement, but noted thermodynamic
discrepancies, like an elevated energy and a depressed density as
compared to Ewald simulations.

Here we use LMF theory along with long-wavelength constraints on the
behavior of charge correlation functions implied by exact expressions
for the dielectric constant in neutral
systems~\cite{Chandler.1977.Dielectric-constant-and-related-equilibrium-properties-of-molecular}
and the related Stillinger-Lovett moment
conditions~\cite{StillingerLovett.1968.General-Restriction-on-Distribution-of-Ions-in-Electrolytes}
for ionic systems to derive analytic expressions for $U_1$ and $P_1$.
Results incorporating this correction for the energy of SPC/E water are also given in
\fref{fig:BulkECorr} and their high accuracy is evident.

\section{Local Molecular Field Theory for Site-Site Molecular
  Simulations \label{sxn:lmf}}
Local molecular field theory provide a general theoretical framework for
assessing and correcting spherical truncations of Coulomb interactions.  The
derivation of \lmf\ theory for systems with Coulomb interactions has
been recently reviewed
elsewhere~\cite{RodgersWeeks.2008.Local-molecular-field-theory-for-the-treatment},
and we will be brief in our discussion here.

 \lmf\ theory divides the $1/r$ potential into short- and
long-ranged parts characterized by the length $\sigma$ as
\begin{equation}
  \frac{1}{r} = \vs + \vl = \frac{\erfc(r/\sigma)}{r} + \frac{\erf(r/\sigma)}{r}.
  \label{eqn:PotSplit}
\end{equation}
This potential separation isolates strong short-ranged and rapidly-varying
Coulomb interactions in $v_0(r)$, while the remaining slowly-varying long-ranged
forces are contained in $v_1(r)$. $v_1(r)$ is proportional
to the electrostatic potential arising from a smooth normalized Gaussian charge distribution with
width $\sigma$, and is defined by the convolution
\begin{equation}
  \label{eqn:v1def}
 v_1(r) \equiv  \frac{1}{\pi^{3/2}\sigma^3}\int e^{-r^{\prime 2}/\sigma^2}\frac{1}{\left| \vect{r} - \vect{r^\prime} \right|} \, d\vect{r}^\prime.
\end{equation}
By construction, $v_{1}(r)$ is slowly-varying in $r$-space over
the \emph{smoothing length} $\sigma$
(see, $e.g.$, Fig.\ 1 in Ref.\ \onlinecite{RodgersWeeks.2008.Local-molecular-field-theory-for-the-treatment}), and contains only
small wave vectors in reciprocal space, as can be seen from its Fourier transform
\begin{equation}
  \label{eqn:v1hat}
\hat{v}_{1}(k) = \frac{4\pi} {k^{2}} \exp [{-(k\sigma )^{2}/4}].
\end{equation}
The short-ranged $v_0(r) \equiv 1/r - v_1(r)$ is then the screened potential resulting
from a point charge surrounded by a neutralizing Gaussian charge
distribution whose width $\sigma$ also sets the scale for the smooth truncation of $v_{0}$. At distances much less than $\sigma$ the force from $v_0(r)$ approaches that from the full $1/r$ potential.

Starting from the exact Yvon-Born-Green
hierarchy~\cite{HansenMcDonald.2006.Theory-of-Simple-Liquids} and
exploiting the slowly-varying nature of \vl, LMF theory
accounts for the averaged effects of the long-ranged
component \vl\ in a mean-field sense by a
rescaled, self-consistent, mean electrostatic potential
$\Vr(\vect{r})$. The short-ranged \vs, with $\sigma$ chosen large
enough to capture relevant nearest-neighbor interactions like core
repulsions and hydrogen bonding, is the spherical truncation used in
\lmf\ theory.  A system in the presence of $\Vr(\vect{r})$ with $1/r$
replaced by the short-ranged \vs\ is referred to as a \emph{mimic
system}, and densities associated with such a system are indicated
by $\rho_R(\vect{r})$.

Often for uniform systems, $\Vr(\vect{r})$ has negligible effect on short-ranged pair 
correlations~\cite{ChenKaurWeeks.2004.Connecting-systems-with-short-and-long,HuRodgersWeeks.2008.On-the-efficient-short-ranged-simulations-of-polar-molecular}.
For such uniform systems, we may simulate simply using \vs\ with
$\Vr(\vect{r})=0$ and thus generate densities $\rho_0(\vect{r})$.
We call this approximation to the full LMF theory the \emph{strong-coupling approximation}
and refer to the resulting truncated water model as \emph{Gaussian-truncated water}.
While not generally true, \lmf\ theory in the strong-coupling approximation is related
to other spherical truncations such as site-site reaction
field~\cite{HummerSoumpasisNeumann.1994.Computer-simulation-of-aqueous-Na-Cl-Electrolytes}
and Wolf
truncations~\cite{WolfKeblinskiPhillpot.1999.Exact-method-for-the-simulation-of-Coulombic-systems}.

\lmf\ theory seeks to obtain the properties of the full, uniform
system from the simulation of the short-ranged system, whose total
energy $U_0$ for Gaussian-truncated SPC/E water includes all
contributions from the Lennard-Jones interactions as well as the
short-ranged components of the Coulomb interactions due to \vs.  Other
authors have used the \vs\ truncation and proposed
numerical corrections to the energy and pressure of ionic systems
based on integral equation methods~\cite{Linse}, but the simple and accurate
analytical corrections possible using moment conditions and our choice of \vs\ and \vl\
as described below in Sections \ref{sxn:energy} and \ref{sxn:press} have not been
previously derived.

\section{Simulation Details \label{sxn:sim}}
The main result of this paper is a general derivation of simple analytical
corrections for the spherical truncation of Coulomb interactions in
simple charged and uncharged site-site molecular models.  In
order to demonstrate the accuracy of these corrections, we also have carried out
a series of simulations of a molecular water model at ambient
conditions.

\begin{figure}[tb]
  \includegraphics[width=2.5in]{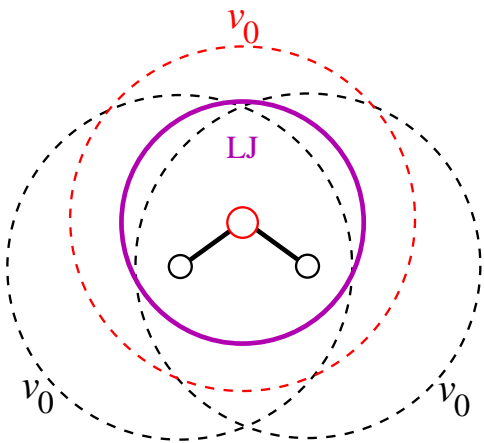}
  \caption{Diagram of Gaussian-truncated \spce\ water.  The
    traditional \spce\ model of water is shown in the center with three
    point charges and a Lennard-Jones core to represent the excluded
    volume.  Gaussian-truncated water is constructed by replacing the
    three point charges with the corresponding short-ranged \vs.}
  \label{fig:water}
\end{figure}

The water model we choose is \spce\
water~\cite{BerendsenGrigeraStraatsma.1987.The-missing-term-in-effective-pair-potentials},
shown in the center of \fref{fig:water}.  A
Lennard-Jones core, depicted by the solid circle with diameter
$\sigma_{LJ}=3.161$~\AA\ accounts for the excluded volume of the molecules,
and point charges are present at each of the atomic sites in order to
represent the charge separation along the OH bonds and to allow for
hydrogen bonding between molecules.  In order to simulate
Gaussian-truncated water, we replace the $1/r$ interaction from
each of these point charges
by the short-ranged \vs\, as represented by the
dashed circles drawn here to scale with diameter $\sigma = 4.5$~\AA .

We carried out a molecular dynamics simulation of a uniform system of
1728 \spce\ water molecules using a modified version of {\sc
  dlpoly2.16}~\cite{Smith.2006.DLPOLY-applications-to-molecular-simulation-II}.
The Berendsen
thermostat~\cite{BerendsenPostmaGunsteren.1984.Molecular-dynamics-with-coupling-to-an-external-bath}
with a relaxation time constant of 0.5~ps is used to maintain the
temperature at 300~K, and, for the final set of data presented, a
Berendsen barostat maintains the pressure.  All simulations use a
timestep of 1~fs.  For the spherical truncations, \sig\ ranges from
3.0~\AA\ to 6.0~\AA, with the cutoff radius ranging from 9.5~\AA~(the
cutoff radius for the Lennard-Jones core) to 13.5~\AA.  As explored in
Refs.~\onlinecite{RodgersWeeks.2008.Interplay-of-local-hydrogen-bonding-and-long-ranged-dipolar}
and~\onlinecite{HuRodgersWeeks.2008.On-the-efficient-short-ranged-simulations-of-polar-molecular},
any \sig\ of 3.0~\AA\ or greater reproduces the short-ranged pair
correlations predicted by Ewald sums quite well.  For our benchmark,
we compare to simulations using three-dimensional Ewald sums with
$\alpha = 0.30$~\AA$^{-1}$ and $k_{\rm max}=10$.  The systems were
each equilibrated for a total of 500 ps, with 1.5 ns of data
collection; error bars were based on 100 ps blocks of data.

\section{Analytical Energy Correction via Moment
  Conditions \label{sxn:energy}}
We assume here that charged interactions arise only between charges
on different molecules, as is the case for typical molecular liquid models like SPC/E water.
These ideas can be extended to larger
molecular species with intramolecular charge-charge interactions
between further-neighbor sites, as briefly discussed in~\aref{sxn:rhoqq}.

The total Coulomb energy $U^q$
for the full system can then be exactly expressed in terms of a two-point
intermolecular charge-density function $\rho^{qq}$ (with units charge$^2$/volume$^2$) as
\begin{align}
  U^q &= \frac{1}{2} \int d\vect{r} \int d\vect{r}^\prime \frac{\rho^{qq}(\vect{r},\vect{r^\prime})}{\vdiff{r}{r^\prime}} \nonumber \\ 
  & = \frac{V}{2}  \int d\vect{r} \, \rho^{qq}(r) \vs + \frac{V}{2}  \int d\vect{r} \, \rho^{qq}(r) \vl ,
  \label{eqn:FullEnergy}
\end{align}
where we have used the uniformity of the fluid and Eq.\ (\ref{eqn:PotSplit}) in the second equality.
The composite function $\rho^{qq}(r)$ is a
charge-weighted linear combination of all intermolecular, two-point
site-site distribution functions
\cite{ChandlerPratt.1976.Statistical-Mechanics-of-Chemical-Equilibria-and-Intramolecular,Chandler.1977.Dielectric-constant-and-related-equilibrium-properties-of-molecular}, and
a detailed expression is given in \aref{sxn:rhoqq}.

As in
Ref.~\onlinecite{DenesyukWeeks.2008.A-new-approach-for-efficient-simulation-of-Coulomb-interactions},
we argue that the first term on the right in Eq.\ (\ref{eqn:FullEnergy})
can be accurately approximated by $U_0^q$, the energy obtained directly from
the Gaussian-truncated water simulation using \vs\ alone, because at
short distances where \vs\ is non-negligible, $\rho_{0}^{qq}(r)$
closely resembles the exact $\rho^{qq}(r)$.  Thus we have
\begin{align}
   \frac{U^{q}}{V} &\approx  \frac{1}{2}  \int d\vect{r} \, \rho_{0}^{qq}(r) \vs + \frac{1}{2}  \int d\vect{r} \, \rho^{qq}(r) \vl.
   \label{eqn:UqFirstApprox}
\end{align}
However, as noted in
Ref.~\onlinecite{ChenKaurWeeks.2004.Connecting-systems-with-short-and-long},
a similar approximation for the second term will fail because \vl\
mainly contains small-wavevector components, exactly the range of
$k$-components where $\rho_{0}^{qq}(r)$ will not accurately represent
the $\rho^{qq}(r)$ of the full system. In fact, this integral will
\emph{diverge} if constraints due to neutrality in ionic systems are
not obeyed. Similar considerations are true for the mixed molecular
systems considered here.

Thus, we again follow the more fruitful path of writing the second term in
$k$-space and approximating the long wavelength behavior of the charge
energy function based on exact relations. For a uniform system, we may
use Parseval's relation and  Eq.\ (\ref{eqn:v1hat})  to reexpress \eref{eqn:UqFirstApprox} exactly as 
\begin{equation}
  \frac{U^{q}}{V} \approx U_0^q + \frac{1}{2} \frac{1}{(2\pi)^3} \int d\vect{k} \frac{4\pi}{k^2} \hat{\rho}^{qq}(k) e^{ -k^2 \sigma^2/4}.
  \label{eqn:U1kspace}
\end{equation}

This choice in Eq.\ (\ref{eqn:v1hat}) of \vl\ in \lmf\ theory allows us to make a highly
useful approximation that highlights its advantages over other
possible potential separations. The Gaussian from $\hat{v}_1(k)$ damps out the large
$k$-contributions to the second term, $U^{q}_1$. Thus for
sufficiently large $\sigma$ we can simply represent $\hat{\rho}^{qq}(k)$
by its two smallest moments in $k$-space, 
\begin{equation}
  \hat{\rho}^{qq}(k) \approx \hat{\rho}^{(0)qq} + \hat{\rho}^{(2)qq} k^2 + \mathcal{O}(k^4).
\end{equation}

We show in \aref{sxn:rhoqq} that $\hat{\rho}^{qq}(k)$ is very simply
related to the basic charge-charge linear response function
$\hat{\chi}^{qq}(k)$ that appears in Chandler's
formula~\cite{Chandler.1977.Dielectric-constant-and-related-equilibrium-properties-of-molecular}
for the dielectric constant of a neutral molecular mixture. We have
generalized the derivation to include both neutral and charged
molecular species in \aref{sxn:rhoqq} and demonstrate the simpler
expansion of $\hat{\chi}^{qq}(k)$ in \aref{sxn:chiqq}. This expression
is a consequence of Stillinger-Lovett-like sum
rules~\cite{HansenMcDonald.2006.Theory-of-Simple-Liquids,Martin.1988.Sum-rules-in-charged-fluids}
arising from the assumption that the potential induced by a test
charge $Q$ in a uniform molecular fluid approaches $4\pi Q/\epsilon
k^2$ at small $k$ to linear order in $Q$. This allows us to relate the
moments of $\hat{\rho}^{qq}$ to the dielectric constant $\epsilon$ and
other molecular properties. A system with mobile ions exhibits
complete screening with $\epsilon = \infty$.

Here we simply state the final expansion of the two-point
charge-density function up to second order in $k$, as derived in
\aref{sxn:rhoqq}.  We find for a general mixture of charged (C) and
neutral (N) polarizable site-site molecules without intramolecular
charge-charge interactions,
\begin{align}
\hat{\rho}^{qq}(k) & =  - \sum_C \rho_C q_C^2 + k^2  \frac{\kT}{4\pi} \frac{\epsilon - 1}{\epsilon} \nonumber \\
& \quad - k^2\sum_N \rho_N \left\{ \frac{1}{3} \mu_N^2 + \kT \alpha_N \right\} \nonumber \\
& \quad + \frac{1}{6} k^2 \sum_{C} \rho_C \sum_{\alpha, \gamma}  q_{\alpha C}q_{\gamma C}\left< l_{\alpha \gamma C}^2 \right>
+ \mathcal{O}(k^4).
\label{eqn:rhomix}
\end{align}
Here, $\rho_C$ and $\rho_N$ are charged and neutral species bulk
densities, $\mu_N$ indicates the dipole moment of a neutral molecule,
and $\alpha_N$ is the molecular polarizability. The final term sums
over the the average of the square of given bond lengths $l_{\alpha
  \gamma C}$ in a charged molecule.  For larger {\sc charmm}- or {\sc
  amber}-like molecular models, a generalization of this approach
leading to related moment-like conditions is possible.

Using this small-moment expansion in \eref{eqn:U1kspace} and noting
that the integrals of Gaussians involved can be analytically
evaluated, we find
\begin{align}
  \frac{U^{q}_1}{V} &\approx - \frac{1}{\sigma \sqrt{\pi}} \sum_C \rho_C q_C^2 +
  \frac{2}{\sigma^3 \sqrt{\pi} }\frac{\kT}{4\pi} \frac{\epsilon -
    1}{\epsilon} \nonumber \\
 & \qquad \qquad - \frac{2}{\sigma^3 \sqrt{\pi}}\sum_N \rho_N \left\{
    \frac{1}{3} \mu_N^2 + \kT \alpha_N \right\} \nonumber \\
  & \qquad \qquad + \frac{1}{3\sigma^3\sqrt{\pi}} \sum_{C}
  \rho_C \sum_{\alpha, \gamma} q_{\alpha C}q_{\gamma C}\left< l_{\alpha \gamma C}^2 \right>.
  \label{eqn:U1correct}
\end{align}

In particular, for bulk SPC/E water, which is neutral and nonpolarizable,
we have
\begin{equation}
  \frac{U^{q}_1}{N} \approx \frac{2 }{\sigma^3 \sqrt{\pi}}  \left( \frac{\kT}{4\pi \rho_w} \frac{\epsilon_w - 1}{\epsilon_w} - \frac{\mu_w^2}{3} \right),
\end{equation}
where $\rho_w$ is the bulk density, $\epsilon_w$ is the dielectric
constant, and $\mu_w$ is the dipole moment of \spce\ water. In
contrast to the expression developed for ionic
solutions~\cite{ChenKaurWeeks.2004.Connecting-systems-with-short-and-long,DenesyukWeeks.2008.A-new-approach-for-efficient-simulation-of-Coulomb-interactions},
the energy correction now incorporates a significant, nontrivial contribution from
the dipole moment. 

In fact $U^{q}_1/N$ is \emph{negative} and may be bounded from above
as $U^{q}_1/N \leq -118.3$~$\frac{{\rm kJ}}{{\rm mol}}\cdot$\AA$^3 /
\sigma^3$ for $T=300$~K, by assuming $\epsilon \rightarrow \infty$ and
using $\mu_w$ determined from the rigid geometry of the \spce\ water
molecule.  In obtaining this numerical expression, recalling that
\eref{eqn:U1correct} was derived using cgs units is crucial.  Since
water has a large dielectric constant and the dipole moment
contribution is large in magnitude, this is actually a very
tight upper bound.  If instead we use the experimental value of
$\epsilon_w=78$, we find $U^{q}_1/N = -118.4$~$\frac{{\rm kJ}}{{\rm
    mol}}\cdot$\AA$^3 / \sigma^3$ for $T=300$~K with variation lying
within error bars of the simulation calculation of $U_0$. 
In \fref{fig:BulkECorr} we used the infinite dielectric constant in our calculation of the
energy correction.

As seen in \fref{fig:BulkECorr}, the inclusion of this correction
brings all of the energies from Gaussian-truncated simulations much
closer to the Ewald energy, shown as a horizontal line. All energies
now lie well within 1\% deviation from the Ewald energy, some with
substantially less error than that, whereas only the three larger
$\sigma$-values without $U^{q}_1$ would lie within the less stringent
5\% deviation suggested as sufficient in
Ref.~\onlinecite{ZahnSchillingKast.2002.Enhancement-of-the-wolf-damped-Coulomb-potential:}.

For solutions of charged particles, previous researchers
obtained similar correction terms for
energies, though their physical basis was less
transparent~\cite{HummerPrattGarcia.1996.Free-energy-of-ionic-hydration,WolfKeblinskiPhillpot.1999.Exact-method-for-the-simulation-of-Coulombic-systems}.
The corrections by Hummer and coworkers relied on an analogy
with the self-interaction in Ewald summations. The corrections by Wolf
and coworkers drew upon the known limiting behaviors for charged
fluids based on the Stillinger-Lovett moment conditions. But the
necessary extension to mixed charged and polar molecular systems was
not appreciated. Combining thermodynamic perturbation theory with an
examination of moment conditions for molecules, as in this paper,
clarifies the general principles involved, and immediately leads to
substantially improved energetics with a simple analytical energy correction.

\section{Analytical Pressure Correction for NVT and NPT
  Simulations \label{sxn:press}}

Deriving a similar correction for the pressure may seem more problematic,
since the pressure cannot be exactly
expressed using only site-site distribution functions~\cite{pressurecorrect},
and to our knowledge no analytic pressure corrections have ever been suggested.
However, LMF theory
provides a general perspective that allows us to arrive at simple pressure
corrections as well. To that end, we express the pressure
thermodynamically as
\begin{equation}
\label{eqn:Pthermo}
  P =  T \left( \frac{\partial S}{\partial V} \right)_{T,\left\{N_M\right\}} - \left( \frac{\partial U}{\partial V} \right)_{T,\left\{N_M\right\}}.
\end{equation}

Since our Gaussian-truncated system with purely short-ranged
interactions reasonably captures the local order and structural
variations expected to dominate the entropy, we expect that $S \approx S_0$
to a very good approximation,
and we use this in the first term on the right in Eq.\ (\ref{eqn:Pthermo}). Corrections from
the long-ranged part of the Coulomb interactions to the pressure
$P_{0}$ obtained directly from the truncated model simulation then
arise from the second term and are simply related to a partial
derivative of $U^{q}_1$ with respect to
volume. 

Using \eref{eqn:U1correct} to express $U_1^q$ in terms of
$\left\{ N_M \right\}$, $V$, and $T$, we find for \emph{rigid}
molecules that only the contribution due to dielectric shielding
depends on volume. Therefore, regardless of the site composition of
the rigid species, we have
\begin{equation}
  P^q_1 = - \left(\frac{\partial U^{q}_1}{\partial V}\right)_{T,\left\{N_M\right\}} = - \frac{\kT}{2 \pi^{3/2} \sigma^3}\frac{\epsilon-1}{\epsilon}.
\end{equation}

\begin{figure}[tb]
  \centering
  \includegraphics[width=3.25in]{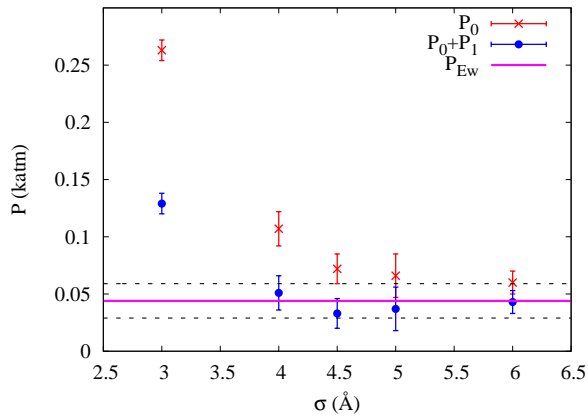}
  \caption{Plot of pressure without ($P_0$ shown in red crosses) and with the long-range correction ($P_0+P_1$ shown in blue circles) for the full range of truncation scales $\sigma$ studied.   Error bars are shown for the data points, and error bars on the Ewald pressure are indicated by the thin dashed lines above and below the thick horizontal line at 0.044 katm.}
  \label{fig:BulkPCorr}
\end{figure}

This correction term is purely negative, just as we can deduce from
our simulated $P_0$ for water shown in \fref{fig:BulkPCorr}.  Using
the experimental dielectric constant of water, $\epsilon_{w} =78$, we
find $P^q_1 = - 3.624$~katm$\cdot$\AA$^3 / \sig^3$.  As shown in
\fref{fig:BulkPCorr}, including $P^q_1$ brings nearly all pressures
into agreement with the Ewald result.  For flexible molecules,
$\mu_N^2$, $\left< l^2\right>$, or $\alpha_N$ could have a volume
dependence leading to a contribution to $P^q_1$. However, for dense
and relatively incompressible systems, such a contribution is likely
quite small.

\begin{figure}[tb]
  \centering
  \includegraphics[width=3.25in]{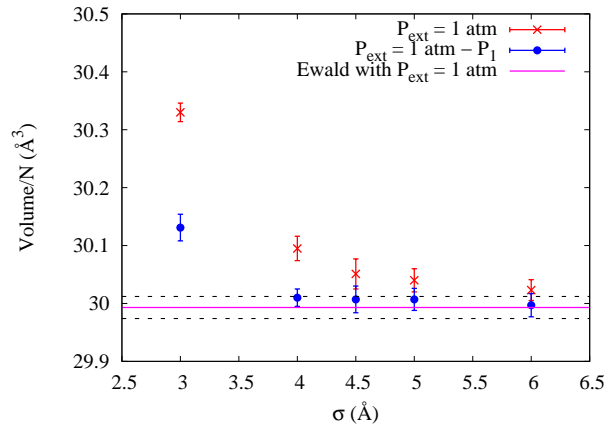}
  \caption{Plots of the volume per particle determined from NPT
    simulation using \vs.  Error bars are as in \fref{fig:BulkPCorr}.
    Applying the $P^q_1$ correction as an external pressure brings
    particle volumes in the mimic system into better agreement with
    the Ewald result.}
  \label{fig:BulkVCorr}
\end{figure}

This analytical pressure correction also proves useful for NPT
simulations of the Gaussian-truncated water. Shown in
\fref{fig:BulkVCorr} is the volume per particle calculated during NPT
simulations carried out at 300 K and 1 atm. The spherically-truncated
water simulations with $P=1$~atm have a higher volume per particle
than the Ewald results as found in
Ref.~\onlinecite{IzvekovSwansonVoth.2008.Coarse-graining-in-interaction-space:-A-systematic-approach}.
However when Gaussian-truncated water is simulated with a corrected
external pressure adjusted to be $P_{\rm ext} = P_0 = 1~{\rm
  atm}-P^q_1(T,\sigma)$, the average volume per particle agrees quite
well with Ewald results for all \sig\ but the smallest of 3.0~\AA. The
latter discrepency simply indicates that the second order $k$-space
expansion for $\hat{\rho}^{qq}$ is insufficient for the smallest \sig\
used.

\section{Concluding Remarks \label{sxn:conc}}

In general, as has been well
established~\cite{DenesyukWeeks.2008.A-new-approach-for-efficient-simulation-of-Coulomb-interactions,ChenKaurWeeks.2004.Connecting-systems-with-short-and-long},
despite the highly accurate local structures obtained when using
reasonable spherical truncations, the impact of the long-ranged forces
on thermodynamics cannot be neglected. We have shown here that high
accuracy is possible for energy, pressure, and density in
spherically-truncated simulations of bulk molecular fluids solely by
using simple, analytical corrections.

Thermodynamic corrections for nonuniform systems treated via \lmf\
theory will be less straightforward. For example, for many of the slab
systems we have simulated, a self-consistent \Vr\ is crucial for
the structure~\cite{RodgersKaurChen.2006.Attraction-between-like-charged-walls:-Short-ranged,RodgersWeeks.2008.Interplay-of-local-hydrogen-bonding-and-long-ranged-dipolar},
and the full LMF theory should be used for the thermodynamics as well.
Corrections to thermodynamics from a strong-coupling
simulation could perhaps be found in some cases
based on the Carnie-Chan sum rules~\cite{Martin.1988.Sum-rules-in-charged-fluids},
a nonuniform analog of the Stillinger-Lovett moment conditions, but
further theoretical development is necessary.
However the simple analytical corrections presented herein should be
immediately useful in correcting the thermodynamics of many bulk
systems of interest.

This work was supported by NSF grants CHE0517818 and CHE0848574. JMR acknowledges the
support of the University of Maryland Chemical Physics fellowship.

\appendix

\section{Derivation of Zeroth and Second Moment Conditions for a Mixture of Neutral and Charged Site-Site Molecules \label{sxn:rhoqq}}

In this Appendix we define the two-point intermolecular charge-density function
$\rho^{qq}(\vect{r}, \vect{r^\prime})$ used
to determine the total Coulomb energy and connect its behavior at small wave vectors
to that of the fundamental charge-charge linear response function used
in the theory of the dielectric constant, as discussed in Appendix B.
These results are used in the
main body to derive corrections to the thermodynamics of uniform
site-site molecular models simulated with spherically truncated
Coulomb interactions.

The total Coulomb energy obtained during simulation of
a small site-site molecular species without any intramolecular charge-charge interactions is
\begin{align}
  U^q &= \left< \frac{1}{2} \sum_{M} \sum_{M^\prime} \sum_{i=1}^{N_M}
    \sum_{j=1}^{N_{M^\prime}} \left(1 - \delta_{MM^\prime}\delta_{ij}
    \right) \right . \nonumber \\
    & \qquad \qquad \times \left. \sum_{\alpha=1}^{n_M} \sum_{\gamma=1}^{n_{M^\prime}}
    \frac{q_{\alpha M} q_{\gamma M^\prime}}{\left| \vect{r}_{i
          M}^{(\alpha)} - \vect{r^\prime}_{jM^\prime}^{(\gamma)} \right|}\right>.
  \label{eqn:energysum}
\end{align}
In this notation, the angular brackets indicate a normalized ensemble
average, $M$ and $M^\prime$ indicate a given molecular species, $i$
and $j$ indicate a given molecule of a given species, and $\alpha$ and
$\gamma$ represent the intramolecular
sites~\cite{ChandlerPratt.1976.Statistical-Mechanics-of-Chemical-Equilibria-and-Intramolecular,Chandler.1977.Dielectric-constant-and-related-equilibrium-properties-of-molecular}.
The Kronecker deltas are necessary to exclude any charge-charge interactions between intramolecular sites
within a given molecule.  This energy $U^q$ can be more compactly represented as
\begin{equation}
  U^q = \frac{1}{2} \int d\vect{r} \int d\vect{r^\prime}
  \frac{\rho^{qq}(\vect{r}, \vect{r^\prime})}{\vdiff{r}{r^\prime}},
  \label{eqn:rhoqq}
\end{equation}
where $\rho^{qq}$ is a two-point intermolecular charge-density function that
explicity excludes any purely intramolecular charge correlations, as
implied by \eref{eqn:energysum} and detailed below.

Comparing Eqs.\ (\ref{eqn:energysum}) and (\ref{eqn:rhoqq}),
we see the composite function $\rho^{qq}(\vect{r},\vect{r}^\prime)$ is a
charge-weighted linear combination of all intermolecular, two-point
site-site distribution functions~\cite{ChandlerPratt.1976.Statistical-Mechanics-of-Chemical-Equilibria-and-Intramolecular,Chandler.1977.Dielectric-constant-and-related-equilibrium-properties-of-molecular}:
\begin{equation}
  \rho^{qq}(\vect{r},\vect{r}^\prime) \equiv \sum_{\alpha M} \sum_{\gamma M^\prime} q_{\alpha M}
  q_{\gamma M^\prime} \rho_{\alpha M \gamma M^\prime} (\vect{r},\vect{r}^\prime).
\end{equation}
For our purposes here, it is more useful to relate this function to the basic charge-charge linear response function used in the theory of the dielectric constant.

For solutions of primitive model ions, Stillinger and Lovett showed
that charge neutrality and screening
place specific requirements on the behavior of the charge density in
$k$-space at small wave vectors~\cite{StillingerLovett.1968.General-Restriction-on-Distribution-of-Ions-in-Electrolytes}.
More generally, for a fluid composed of charged and polar molecules,
the dielectric screening behavior of the molecules places restrictions
on the decay of the two-point charge density $\rho^{qq}$. Based on this
observation, we are able to harness a theoretical development of
Chandler~\cite{Chandler.1977.Dielectric-constant-and-related-equilibrium-properties-of-molecular}
that expresses the dielectric constant $\epsilon$ of polar molecules in terms of an exact sum of
charge-density-weighted site pair correlation functions. We generalize the
derivation to include both charged and neutral site-site molecules and
we take the dielectric constant as a given. From this vantage point,
we may instead use these relations to place requirements on the decay
of the two-point charge density $\rho^{qq}$.

We first define the instantaneous single-point total charge-density
$\rho^q(\vect{r},\allpos)$, a function of both a given
external position $\vect{r}$ and the set
$\allpos \equiv \left\{ \vect{R}_{iM} \right\} \equiv \left\{ \vect{r}_{iM}^{(\alpha)} \right\} $
of positions of all mobile charged sites in a given configuration, as
\begin{equation}
  \rho^q (\vect{r},\allpos) \equiv \sum_{M} \sum_{i=1}^{N_M}
  \sum_{\alpha=1}^{n_M} q_{\alpha M} \, \delta ( \vect{r} - \vect{r}_{iM}^{(\alpha)} ).
  \label{eqn:rhoqdef}
\end{equation}
With such a definition, the ensemble-averaged charge density profile $\rho^q(\vect{r})$ is
\begin{equation}
  \rho^q(\vect{r}) = \left< \rho^q (\vect{r},\allpos) \right>.
\end{equation}
In the case of a uniform system, $\rho^q(\vect{r}) = 0$.

Comparing Eqs.\ (\ref{eqn:energysum}) and (\ref{eqn:rhoqq}) and using Eq.\ (\ref{eqn:rhoqdef}),
we may also express the two point charge function $\rho^{qq}(\vdiff{r}{r^\prime})$ for a
uniform system as
\begin{align}
&\rho^{qq}(\vdiff{r}{r^\prime}) = \left< \rho^q (\vect{r},\allpos)
  \rho^q (\vect{r^\prime},\allpos)\right> \nonumber \\
& \qquad- \left< \sum_{M} N_M \sum_{\alpha=1}^{n_M}
  \sum_{\gamma=1}^{n_M} q_{\alpha M} q_{\gamma M} \delta ( \vect{r}-\vect{r}_{1M}^{(\alpha)}) \delta ( \vect{r^\prime}-\vect{r}_{1M}^{(\gamma)} )\right>.
  \label{eqn:rhoqquniform}
\end{align}
We have used the equivalence of all molecules of type $M$ in the last term.
This term removes purely intramolecular charge-density correlations;
we shall determine the small-$k$ contributions from this term based on
well-known molecular properties using the approach of
Chandler~\cite{Chandler.1977.Dielectric-constant-and-related-equilibrium-properties-of-molecular}
later in this appendix.

The first term, in contrast, is exactly the
charge-charge linear response function for a uniform neutral system:
\begin{align}
  \left< \rho^q(\vect{r},\allpos) \rho^q
    (\vect{r^\prime},\allpos) \right> &= \left<
    \delta\rho^q(\vect{r},\allpos) \delta \rho^q
    (\vect{r^\prime},\allpos) \right> \nonumber \\
  &=  \chi^{qq}(\vdiff{r}{r^\prime}).
  \label{eqn:chiqquniform}
\end{align}
Here $\delta \rho^q(\vect{r},\allpos) \equiv \rho^q(\vect{r},\allpos)
- \left< \rho^q (\vect{r},\allpos) \right>$.  Physically $\chi^{qq}$
describes the coupling between charge-density fluctuations at
positions $\vect{r}$ and $\vect{r^\prime}$. As is well
established~\cite{Chandler.1977.Dielectric-constant-and-related-equilibrium-properties-of-molecular,HansenMcDonald.2006.Theory-of-Simple-Liquids},
such a function is intimately related to the dielectric behavior of
the fluid at long distances, and furthermore, may be easily analyzed
based on basic electrostatics and standard definitions of the
functional derivative.  This relationship is discussed in
\aref{sxn:chiqq}.

Our goal is to write a small-$k$ expansion of the two-point intermolecular  charge density,
\begin{equation}
\hat{\rho}^{qq}(k) \approx \hat{\rho}^{(0)qq} + k^2 \hat{\rho}^{(2)qq} + \mathcal{O}(k^4),
\end{equation}
where from \erefs{eqn:rhoqquniform}{eqn:chiqquniform}
\begin{align}
& \rho^{qq}(\vdiff{r}{r^\prime}) = \chi^{qq}\left( \vdiff{r}{r^\prime} \right) \nonumber \\
& \qquad - \left< \sum_{M} N_M \sum_{\alpha=1}^{n_M}
  \sum_{\gamma=1}^{n_M} q_{\alpha M} q_{\gamma M} \delta ( \vect{r}-\vect{r}_{1M}^{(\alpha)}
  ) \delta ( \vect{r^\prime}-\vect{r}_{1M}^{(\gamma)} )\right>.
  \label{eqn:rhochi}
\end{align}
As shown in \aref{sxn:chiqq}, the
charge-charge linear response function $\hat{\chi}^{qq}$ may be expanded as
\begin{equation}
\hat{\chi}^{qq}(k) = 0 + \frac{\kT}{4\pi} \left( 1 - \frac{1}{\epsilon} \right) k^2  + \mathcal{O}(k^4).
\label{eqn:chiexp}
\end{equation}

Now we must remove the intramolecular contributions as described by the last
term in \eref {eqn:rhochi}. Defining the conditional singlet
intramolecular site density functions $\varrho_{\alpha | \gamma
  M}(\vect{r}| \vect{r^\prime})$ for $\alpha \neq \gamma$ as
\begin{equation}
  \rho_{\gamma M}(\vect{r^\prime}) \varrho_{\alpha | \gamma M}(\vect{r}|\vect{r^\prime}) = \left< {N_M} \delta ( \vect{r}-\vect{r}_{1M}^{(\alpha)}
  ) \delta ( \vect{r^\prime}-\vect{r}_{1M}^{(\gamma)} ) \right>,
\end{equation}
 and applying consequences of uniformity, \eref {eqn:rhochi} can be written as
\begin{align}
&  \rho^{qq}(\vdiff{r}{r^\prime}) = \chi^{qq}\left( \vdiff{r}{r^\prime}
  \right) \nonumber \\
& \qquad - 
  \sum_{M} \rho_M \sum_{\alpha, \gamma} q_{\alpha M} q_{\gamma M}
 \omega_{\alpha \gamma M}(\vdiff{r}{r^\prime}),
\label{eqn:rhoqqunif}
\end{align}
where
\begin{equation}
\omega_{\alpha \gamma M}(\vdiff{r}{r^\prime}) \equiv
 \delta_{\alpha \gamma} \delta(\vect{r}-\vect{r}^\prime) +  \varrho_{\alpha | \gamma M}(\vdiff{r}{r^\prime}).
\end{equation}

For neutral molecules, Chandler demonstrated that the small-$k$
components of $\hat{\omega}_{\alpha \gamma M}(k)$ are related to simple
properties of the molecule.   For both charged and uncharged molecules, the zeroth
moment of $\hat{\omega}_{\alpha \gamma M}$ is simply
\begin{align}
  \hat{\omega}_{\alpha \gamma M} ^{(0)} &= \delta_{\alpha \gamma} + \int d\vect{r} \,\varrho_{\alpha | \gamma M} (\vect{r}) \nonumber \\
  &= \delta_{\alpha \gamma} + (1 - \delta_{\alpha \gamma}) = 1.
\end{align}
Using this exact expression in Eq.\ (\ref{eqn:rhoqqunif}) yields
\begin{equation}
\hat{\rho}^{(0)qq} = \hat{\chi}^{(0)qq} -  \sum_{M} \rho_M
\sum_{\alpha, \gamma} q_{\alpha M} q_{\gamma M} = - \sum_M \rho_M q_M^2,
\end{equation}
an expression encompassing the standard zeroth moment condition for
ions~\cite{StillingerLovett.1968.General-Restriction-on-Distribution-of-Ions-in-Electrolytes} and the
zeroth moment for neutral molecular
species~\cite{Chandler.1977.Dielectric-constant-and-related-equilibrium-properties-of-molecular}.

The expression for $\hat{\omega}^{(2)}_M$ determined by
Chandler~\cite{Chandler.1977.Dielectric-constant-and-related-equilibrium-properties-of-molecular} may be written most generally as
\begin{align}
  \hat{\omega}^{(2)}_M &\equiv \sum_{\alpha \neq \gamma} q_{\alpha M} q_{\gamma M} \hat{\omega}_{\alpha \gamma M}^{(2)} \nonumber \\ 
  &= - \frac{1}{6} \int d\vect{r} \sum_{\alpha \neq \gamma} q_{\alpha M}  q_{\gamma M} \varrho_{\alpha | \gamma M}(\vect{r}) r^2 \nonumber \\
&= -\frac{1}{6} \sum_{\alpha \neq \gamma} q_{\alpha M} q_{\gamma M} \left< l_{\alpha \gamma M}^2 \right>,
\end{align}
where $l_{\alpha \gamma M}$ is the bondlength between sites $\alpha$
and $\gamma$ for a molecule of species $M$. As shown in
Ref.~\onlinecite{Chandler.1977.Dielectric-constant-and-related-equilibrium-properties-of-molecular},
for a \emph{neutral} molecule indicated by $N$ below, the final
summation in the above equation is simply related to the molecular dipole moment
$\mu_N$ and the molecular polarizability $\alpha_N$ as
\begin{equation}
\hat{\omega}^{(2)}_N = \sum_{\alpha \neq \gamma} q_{\alpha N} q_{\gamma N} \hat{\omega}_{\alpha \gamma N}^{(2)} =  \frac{1}{3} \mu_N^2 + \kT \alpha_N.
\end{equation}
This relationship does not hold for a charged molecule since the
dipole moment then depends on the choice of coordinate system.

Distinguishing charged species ($C$) and  neutral species ($N$) where $\{ M \} = \{ N \} \cup \{ C \}$, and
without substituting for $\hat{\omega}_N^{(2)}$ and $\hat{\omega}_C^{(2)}$, we find
\begin{align}
\hat{\rho}^{(2)qq} &= \frac{\kT}{4\pi} \frac{\epsilon -
  1}{\epsilon}- \sum_M \rho_M \sum_{\alpha \neq \gamma} q_{\alpha M}
q_{\gamma M} \hat{\omega}_{\alpha \gamma M}^{(2)} \nonumber \\
&= \frac{\kT}{4\pi} \frac{\epsilon -
  1}{\epsilon} - \sum_N \rho_N \hat{\omega}_N^{(2)} - \sum_C \rho_C \hat{\omega}_C^{(2)}.
\end{align}
Thus, we may write a general expression for $\hat{\rho}^{qq}$ in $k$-space.
Utilizing the expressions for $\hat{\omega}_N^{(2)}$ and $\hat{\omega}_C^{(2)}$, we have
\begin{align}
\hat{\rho}^{qq}(k) = &- \sum_C \rho_C q_C^2 + k^2 \frac{\kT}{4\pi} \frac{\epsilon - 1}{\epsilon} \nonumber \\
& - k^2 \sum_N \rho_N \left\{ \frac{1}{3} \mu_N^2 + \kT \alpha_N \right\} \nonumber \\
& + k^2 \frac{1}{6} \sum_{C} \rho_C \sum_{\alpha \neq \gamma} q_{\alpha C}q_{\gamma C}\left< l_{\alpha \gamma C}^2 \right> + \mathcal{O}\left( k^4 \right).
\end{align}

Unlike the simple expansion of $\hat{\chi}^{qq}(k)$ in
\eref{eqn:chiexp}, we see that the small-$k$ behavior of
$\hat{\rho}^{qq}(k)$ depends on several simple properties of the
solution as a whole, like the dipole moment and polarizability of
individual neutral molecules, and the net molecular charge and the
average square bond lengths of charged molecules, as well as the
dielectric constant. Thus by knowing simple single molecule properties
and the long wavelength dielectric constant, we know how
intermolecular charge-charge correlations decay in solution. This is
the essential idea used to develop energy and pressure corrections for
simulations of bulk liquids using molecular models with truncated
Coulomb interactions.

A related expression may be developed for larger molecular species
with intramolecular charge-charge interactions given by {\sc charmm}-
or {\sc amber}-like molecular models by modifying the total Coulomb
energy to solely exclude the charge-charge interactions of sites $i$ and $j$
within two bonded connections of one another, using a ``bonding function''
$B_{M}(i,j)$ that acts similar to the product of Kronecker deltas in \eref {eqn:energysum}.
In such cases, the
expansion of $\hat{\chi}^{qq}$ remains the same but the remaining
contribution to $\hat{\rho}^{qq}$ may no longer be simply represented
using only whole-molecule properties such as the dipole moment and
polarizability.

\section{Exact Moment Conditions on Charge-Charge Linear Response \label{sxn:chiqq}}

We present the following moment conditions for $\chi^{qq}$ distinct
from the molecular-specific analysis found in \aref{sxn:rhoqq} because
the results are more general than the specific site-site molecules chosen.
The analysis of the behavior of $\hat{\chi}^{qq}$ at small $k$ is similar to
that found in
References~\onlinecite{HansenMcDonald.2006.Theory-of-Simple-Liquids}~and~\onlinecite{Chandler.1977.Dielectric-constant-and-related-equilibrium-properties-of-molecular}
and the final results are identical.  Our derivation is simpler because we
focus directly on the total charge density; this also allows us to
derive both the Stillinger-Lovett moment conditions for charged systems
and the formula for the dielectric constant of a polar mixture from the same footing.

The electrostatic potential at $\vect{r}$ induced by a fixed external
charge distribution $\rho^q_{\rm ext}(\vect{r^\prime})$ (\eg, a test
charge $Q$ placed at the origin, as considered by Chandler~\cite{Chandler.1977.Dielectric-constant-and-related-equilibrium-properties-of-molecular})
is given by
\begin{equation}
  \mathcal{V}_{\rm ext}(\vect{r}) = \int \frac{\rho^q_{\rm ext}(\vect{r^\prime})}{\vdiff{r}{r^\prime}} d \vect{r}^\prime,
\end{equation}
and the associated electrostatic energy for a particular microscopic
configuration characterized by the set of molecular positions $\allpos$ is then
\begin{equation}
  U^q_{\rm ext}(\allpos) = \int \rho^q(\vect{r},\allpos) \mathcal{V}_{\rm ext}(\vect{r}) d \vect{r}.
\end{equation}
Here $\rho^q(\vect{r},\allpos)$ is the total configurational charge density,
defined in the particular case of a mixture of site-site molecules
by Eq.\ (\ref{eqn:rhoqdef}). This energy contribution will appear in the nonuniform
system's Hamiltonian when $\mathcal{V}_{\rm ext}(\vect{r})$ is nonzero.

As such, we know from standard definitions of functional
differentiation of free
energies~\cite{HansenMcDonald.2006.Theory-of-Simple-Liquids,Percus.J.The-pair-distribution-function-in-classical-statistical.1964}
that
\begin{equation}
  \frac{\delta \left[ -\beta A \right]}{\delta \left[ -\beta
      \mathcal{V}_{\rm ext}(\vect{r})\right]} = \left< \rho^q ( \vect{r},
    \allpos)\right>_{\mathcal{V}} \equiv  \rho^q_{\mathcal{V}}(\vect{r}) ,
\end{equation}
where $\beta\equiv(\kT)^{-1}$ and the subscript $\mathcal{V}$ indicates that the ensemble average is taken in the presence of an external potential.
Similarly we have
\begin{align}
  \frac{\delta \rho^q_{\mathcal{V}}(\vect{r})}{\delta \left[ -\beta
      \mathcal{V}_{\rm ext}(\vect{r}^\prime) \right]} &=   \frac{\delta \left[ -\beta A \right] }{\delta \left[ -\beta
      \mathcal{V}_{\rm ext}(\vect{r}) \right] \delta \left[ -\beta
      \mathcal{V}_{\rm ext}(\vect{r}^\prime) \right]} \nonumber \\ 
  &= \chi^{qq}_{\mathcal{V}}\left(
    \vect{r}, \vect{r^\prime}\right).
  \label{eqn:chiv}
\end{align}

The total electrostatic potential at position $\vect{r}$ in the nonuniform fluid is then given by the sum of the external potential and the induced polarization potential:
\begin{align}
  \mathcal{V}_{\rm tot}(\vect{r}) &= \mathcal{V}_{\rm ext}(\vect{r}) +
  \mathcal{V}_{\rm pol}(\vect{r}) \nonumber \\
  &=  \mathcal{V}_{\rm ext}(\vect{r}) + \int d\vect{r}^\prime \,
  \frac{\rho^{q}_{\mathcal{V}}(\vect{r}^\prime)}{\vdiff{r}{r^\prime}}.
\end{align}
To get a formula for the dielectric constant we expand about the
uniform neutral system and evaluate $\mathcal{V}_{\rm pol}$ to linear order in
$\mathcal{V}_{\rm ext}$ using \eref{eqn:chiv}. This gives
\begin{align}
  \mathcal{V}_{\rm pol}(\vect{r}) &\approx \int 
  \frac{d\vect{r}^\prime}{\vdiff{r}{r^\prime}}  \int d\vect{r}^{\prime \prime} \, \frac{\delta
  \rho^{q}_{\mathcal{V}}(\vect{r}^\prime)}{\delta \left[ -\beta
     \mathcal{V}_{\rm ext}(\vect{r}^{\prime \prime}) \right]} \left[ -\beta \mathcal{V}_{\rm
  ext} (\vect{r}^{\prime \prime})\right], \nonumber \\
  &= - \int \,
  \frac{d\vect{r}^\prime}{\vdiff{r}{r^\prime}}  \int  d\vect{r}^{\prime \prime} \, \beta \chi^{qq} \left(
    \vdiff{r^\prime}{r^{\prime \prime}} \right) \mathcal{V}_{\rm
  ext} (\vect{r}^{\prime \prime}).
\end{align}
Here $\chi^{qq}$ is the linear response function in the uniform fluid as in \eref{eqn:chiqquniform}.
Taking the Fourier transform of the final equation, we find
\begin{equation}
  \hat{\mathcal{V}}_{\rm tot}(\vect{k}) = \hat{\mathcal{V}}_{\rm ext}(\vect{k})- \frac{4
    \pi}{k^2} \beta \hat{\chi}^{qq}(k) \hat{\mathcal{V}}_{\rm ext}(\vect{k}).
\end{equation}
Thus to linear order we have
\begin{equation}
 \frac{\hat{\mathcal{V}}_{\rm tot}(\vect{k})}{\hat{\mathcal{V}}_{\rm ext}(\vect{k})} = 1 -
 \frac{4 \pi \beta }{k^2} \hat{\chi}^{qq}(k).
\end{equation}

Phenomenologically, we know that in the limit of $\vect{k} \rightarrow
0$, this ratio of the total electrostatic potential to the
externally-imposed potential is exactly $1/\epsilon$. Therefore, we
find for our molecular mixture the general result
\begin{equation}
  \lim_{\vect{k} \rightarrow 0} \left( 1 - \frac{4 \pi \beta}{k^2}
    \hat{\chi}^{qq}(k) \right) = \frac{1}{\epsilon}.
  \label{eqn:ChiLimit}
\end{equation}
Based on the limit in \eref{eqn:ChiLimit}, and expanding $\hat{\chi}^{qq}$ for small $k$
as $\hat{\chi}^{(0)qq} + \hat{\chi}^{(2)qq}k^2$, we have 
\begin{align}
  \hat{\chi}^{(0)qq} &= 0 \nonumber \\
  4 \pi \beta \hat{\chi}^{(2)qq} &= 1 - \frac{1}{\epsilon}.
  \label{eqn:chi0}
\end{align}
Any mixture with mobile ions acts as a conductor with $\epsilon = \infty$ in Eq.\ (\ref{eqn:chi0}),
independent of the nature of the neutral components, and these results then reduce to the Stillinger-Lovett
moment conditions\cite{StillingerLovett.1968.General-Restriction-on-Distribution-of-Ions-in-Electrolytes}.


\begin{thebibliography}{30}
\expandafter\ifx\csname natexlab\endcsname\relax\def\natexlab#1{#1}\fi
\expandafter\ifx\csname bibnamefont\endcsname\relax
  \def\bibnamefont#1{#1}\fi
\expandafter\ifx\csname bibfnamefont\endcsname\relax
  \def\bibfnamefont#1{#1}\fi
\expandafter\ifx\csname citenamefont\endcsname\relax
  \def\citenamefont#1{#1}\fi
\expandafter\ifx\csname url\endcsname\relax
  \def\url#1{\texttt{#1}}\fi
\expandafter\ifx\csname urlprefix\endcsname\relax\def\urlprefix{URL }\fi
\providecommand{\bibinfo}[2]{#2}
\providecommand{\eprint}[2][]{\url{#2}}

\bibitem[{\citenamefont{Duan et~al.}(2003)\citenamefont{Duan, Wu, Chowdhury,
  Lee, Xiong, Zhang, Yang, Cieplak, Luo, Lee
  et~al.}}]{DuanWuChowdhury.2003.A-point-charge-force-field-for-molecular-mechanics}
\bibinfo{author}{\bibfnamefont{Y.}~\bibnamefont{Duan}},
  \bibinfo{author}{\bibfnamefont{C.}~\bibnamefont{Wu}},
  \bibinfo{author}{\bibfnamefont{S.}~\bibnamefont{Chowdhury}},
  \bibinfo{author}{\bibfnamefont{M.}~\bibnamefont{Lee}},
  \bibinfo{author}{\bibfnamefont{G.}~\bibnamefont{Xiong}},
  \bibinfo{author}{\bibfnamefont{W.}~\bibnamefont{Zhang}},
  \bibinfo{author}{\bibfnamefont{R.}~\bibnamefont{Yang}},
  \bibinfo{author}{\bibfnamefont{P.}~\bibnamefont{Cieplak}},
  \bibinfo{author}{\bibfnamefont{R.}~\bibnamefont{Luo}},
  \bibinfo{author}{\bibfnamefont{T.}~\bibnamefont{Lee}}, \bibnamefont{et~al.},
  \bibinfo{journal}{J. Comp. Chem.} \textbf{\bibinfo{volume}{24}},
  \bibinfo{pages}{1999} (\bibinfo{year}{2003}).

\bibitem[{\citenamefont{MacKerell et~al.}(1998)\citenamefont{MacKerell,
  Bashford, Bellott, R.~L.~Dunbrack, Evanseck, Field, Fischer, Gao, Guo, Ha
  et~al.}}]{MacKerellBashfordBellott.1998.All-Atom-Empirical-Potential-for-Molecular-Modeling}
\bibinfo{author}{\bibfnamefont{J.~A.~D.} \bibnamefont{MacKerell}},
  \bibinfo{author}{\bibfnamefont{D.}~\bibnamefont{Bashford}},
  \bibinfo{author}{\bibfnamefont{M.}~\bibnamefont{Bellott}},
  \bibinfo{author}{\bibfnamefont{J.}~\bibnamefont{R.~L.~Dunbrack}},
  \bibinfo{author}{\bibfnamefont{J.~D.} \bibnamefont{Evanseck}},
  \bibinfo{author}{\bibfnamefont{M.~J.} \bibnamefont{Field}},
  \bibinfo{author}{\bibfnamefont{S.}~\bibnamefont{Fischer}},
  \bibinfo{author}{\bibfnamefont{J.}~\bibnamefont{Gao}},
  \bibinfo{author}{\bibfnamefont{H.}~\bibnamefont{Guo}},
  \bibinfo{author}{\bibfnamefont{S.}~\bibnamefont{Ha}}, \bibnamefont{et~al.},
  \bibinfo{journal}{J. Phys. Chem. B} \textbf{\bibinfo{volume}{102}},
  \bibinfo{pages}{3586} (\bibinfo{year}{1998}).

\bibitem[{\citenamefont{Frenkel and
  Smit}(2002)}]{FrenkelSmit.2002.Understanding-Molecular-Simulation:-From-Algorithms}
\bibinfo{author}{\bibfnamefont{D.}~\bibnamefont{Frenkel}} \bibnamefont{and}
  \bibinfo{author}{\bibfnamefont{B.}~\bibnamefont{Smit}},
  \emph{\bibinfo{title}{Understanding Molecular Simulation: {F}rom Algorithms
  to Applications}} (\bibinfo{publisher}{Academic Press}, \bibinfo{address}{New
  York}, \bibinfo{year}{2002}), \bibinfo{edition}{2nd} ed.

\bibitem[{\citenamefont{Fennell and
  Gezelter}(2006)}]{FennellGezelter.2006.Is-the-Ewald-summation-still-necessary-Pairwise}
\bibinfo{author}{\bibfnamefont{C.}~\bibnamefont{Fennell}} \bibnamefont{and}
  \bibinfo{author}{\bibfnamefont{J.}~\bibnamefont{Gezelter}},
  \bibinfo{journal}{J. Chem. Phys.} \textbf{\bibinfo{volume}{124}},
  \bibinfo{pages}{234104} (\bibinfo{year}{2006}).

\bibitem[{\citenamefont{Wolf et~al.}(1999)\citenamefont{Wolf, Keblinski,
  Phillpot, and
  Eggebrecht}}]{WolfKeblinskiPhillpot.1999.Exact-method-for-the-simulation-of-Coulombic-systems}
\bibinfo{author}{\bibfnamefont{D.}~\bibnamefont{Wolf}},
  \bibinfo{author}{\bibfnamefont{P.}~\bibnamefont{Keblinski}},
  \bibinfo{author}{\bibfnamefont{S.}~\bibnamefont{Phillpot}}, \bibnamefont{and}
  \bibinfo{author}{\bibfnamefont{J.}~\bibnamefont{Eggebrecht}},
  \bibinfo{journal}{J. Chem. Phys.} \textbf{\bibinfo{volume}{110}},
  \bibinfo{pages}{8254} (\bibinfo{year}{1999}).

\bibitem[{\citenamefont{Nezbeda}(2005)}]{Nezbeda.2005.Towards-a-Unified-View-of-Fluids}
\bibinfo{author}{\bibfnamefont{I.}~\bibnamefont{Nezbeda}},
  \bibinfo{journal}{Mol. Phys.} \textbf{\bibinfo{volume}{103}},
  \bibinfo{pages}{59} (\bibinfo{year}{2005}).

\bibitem[{\citenamefont{Izvekov et~al.}(2008)\citenamefont{Izvekov, Swanson,
  and
  Voth}}]{IzvekovSwansonVoth.2008.Coarse-graining-in-interaction-space:-A-systematic-approach}
\bibinfo{author}{\bibfnamefont{S.}~\bibnamefont{Izvekov}},
  \bibinfo{author}{\bibfnamefont{J.~M.~J.} \bibnamefont{Swanson}},
  \bibnamefont{and} \bibinfo{author}{\bibfnamefont{G.~A.} \bibnamefont{Voth}},
  \bibinfo{journal}{J. Phys. Chem. A} \textbf{\bibinfo{volume}{112}},
  \bibinfo{pages}{4711} (\bibinfo{year}{2008}).

\bibitem[{\citenamefont{Hummer et~al.}(1994)\citenamefont{Hummer, Soumpasis,
  and
  Neumann}}]{HummerSoumpasisNeumann.1994.Computer-simulation-of-aqueous-Na-Cl-Electrolytes}
\bibinfo{author}{\bibfnamefont{G.}~\bibnamefont{Hummer}},
  \bibinfo{author}{\bibfnamefont{D.}~\bibnamefont{Soumpasis}},
  \bibnamefont{and} \bibinfo{author}{\bibfnamefont{M.}~\bibnamefont{Neumann}},
  \bibinfo{journal}{J. Phys. -- Condens. Matt.} \textbf{\bibinfo{volume}{6}},
  \bibinfo{pages}{A141} (\bibinfo{year}{1994}).

\bibitem[{\citenamefont{Feller et~al.}(1996)\citenamefont{Feller, Pastor,
  Rojnuckarin, Bogusz, and
  Brooks}}]{FellerPastorRojnuckari.1996.Effect-of-Electrostatic-Force-Truncation-on-Interfacial}
\bibinfo{author}{\bibfnamefont{S.~E.} \bibnamefont{Feller}},
  \bibinfo{author}{\bibfnamefont{R.~W.} \bibnamefont{Pastor}},
  \bibinfo{author}{\bibfnamefont{A.}~\bibnamefont{Rojnuckarin}},
  \bibinfo{author}{\bibfnamefont{S.}~\bibnamefont{Bogusz}}, \bibnamefont{and}
  \bibinfo{author}{\bibfnamefont{B.~R.} \bibnamefont{Brooks}},
  \bibinfo{journal}{J. Phys. Chem.} \textbf{\bibinfo{volume}{100}},
  \bibinfo{pages}{17011} (\bibinfo{year}{1996}).

\bibitem[{\citenamefont{Spohr}(1997)}]{Spohr.1997.Effect-of-Electrostatic-Boundary-Conditions-and-System}
\bibinfo{author}{\bibfnamefont{E.}~\bibnamefont{Spohr}}, \bibinfo{journal}{J.
  Chem. Phys.} \textbf{\bibinfo{volume}{107}}, \bibinfo{pages}{6342}
  (\bibinfo{year}{1997}).

\bibitem[{\citenamefont{Zahn et~al.}(2002)\citenamefont{Zahn, Schilling, and
  Kast}}]{ZahnSchillingKast.2002.Enhancement-of-the-wolf-damped-Coulomb-potential:}
\bibinfo{author}{\bibfnamefont{D.}~\bibnamefont{Zahn}},
  \bibinfo{author}{\bibfnamefont{B.}~\bibnamefont{Schilling}},
  \bibnamefont{and} \bibinfo{author}{\bibfnamefont{S.}~\bibnamefont{Kast}},
  \bibinfo{journal}{J. Phys. Chem. B} \textbf{\bibinfo{volume}{106}},
  \bibinfo{pages}{10725} (\bibinfo{year}{2002}).

\bibitem[{\citenamefont{Weeks et~al.}(1998)\citenamefont{Weeks, Katsov, and
  Vollmayr}}]{WeeksKatsovVollmayr.1998.Roles-of-repulsive-and-attractive-forces-in-determining}
\bibinfo{author}{\bibfnamefont{J.~D.} \bibnamefont{Weeks}},
  \bibinfo{author}{\bibfnamefont{K.}~\bibnamefont{Katsov}}, \bibnamefont{and}
  \bibinfo{author}{\bibfnamefont{K.}~\bibnamefont{Vollmayr}},
  \bibinfo{journal}{Phys. Rev. Lett.} \textbf{\bibinfo{volume}{81}},
  \bibinfo{pages}{4400} (\bibinfo{year}{1998}).

\bibitem[{\citenamefont{Rodgers and
  Weeks}(2008{\natexlab{a}})}]{RodgersWeeks.2008.Local-molecular-field-theory-for-the-treatment}
\bibinfo{author}{\bibfnamefont{J.~M.} \bibnamefont{Rodgers}} \bibnamefont{and}
  \bibinfo{author}{\bibfnamefont{J.~D.} \bibnamefont{Weeks}},
  \bibinfo{journal}{J. Phys. -- Condens. Matt.} \textbf{\bibinfo{volume}{20}},
  \bibinfo{pages}{494206} (\bibinfo{year}{2008}{\natexlab{a}}).

\bibitem[{\citenamefont{Chen and
  Weeks}(2006)}]{ChenWeeks.2006.Local-molecular-field-theory-for-effective}
\bibinfo{author}{\bibfnamefont{Y.~G.} \bibnamefont{Chen}} \bibnamefont{and}
  \bibinfo{author}{\bibfnamefont{J.~D.} \bibnamefont{Weeks}},
  \bibinfo{journal}{Proc. Nat. Acad. Sci. USA} \textbf{\bibinfo{volume}{103}},
  \bibinfo{pages}{7560} (\bibinfo{year}{2006}).

\bibitem[{\citenamefont{Rodgers et~al.}(2006)\citenamefont{Rodgers, Kaur, Chen,
  and
  Weeks}}]{RodgersKaurChen.2006.Attraction-between-like-charged-walls:-Short-ranged}
\bibinfo{author}{\bibfnamefont{J.~M.} \bibnamefont{Rodgers}},
  \bibinfo{author}{\bibfnamefont{C.}~\bibnamefont{Kaur}},
  \bibinfo{author}{\bibfnamefont{Y.-G.} \bibnamefont{Chen}}, \bibnamefont{and}
  \bibinfo{author}{\bibfnamefont{J.~D.} \bibnamefont{Weeks}},
  \bibinfo{journal}{Phys. Rev. Lett.} \textbf{\bibinfo{volume}{97}},
  \bibinfo{pages}{097801} (\bibinfo{year}{2006}).

\bibitem[{\citenamefont{Rodgers and
  Weeks}(2008{\natexlab{b}})}]{RodgersWeeks.2008.Interplay-of-local-hydrogen-bonding-and-long-ranged-dipolar}
\bibinfo{author}{\bibfnamefont{J.~M.} \bibnamefont{Rodgers}} \bibnamefont{and}
  \bibinfo{author}{\bibfnamefont{J.~D.} \bibnamefont{Weeks}},
  \bibinfo{journal}{Proc. Nat. Acad. Sci. USA} \textbf{\bibinfo{volume}{105}},
  \bibinfo{pages}{19136 } (\bibinfo{year}{2008}{\natexlab{b}}).

\bibitem[{\citenamefont{Rodgers and Weeks}(in
  preparation)}]{RodgersWeeks.2008.Water-in-strong-electric-fields:-Insights}
\bibinfo{author}{\bibfnamefont{J.~M.} \bibnamefont{Rodgers}} \bibnamefont{and}
  \bibinfo{author}{\bibfnamefont{J.~D.} \bibnamefont{Weeks}} (\bibinfo{year}{in
  preparation}).

\bibitem[{\citenamefont{Chen et~al.}(2004)\citenamefont{Chen, Kaur, and
  Weeks}}]{ChenKaurWeeks.2004.Connecting-systems-with-short-and-long}
\bibinfo{author}{\bibfnamefont{Y.~G.} \bibnamefont{Chen}},
  \bibinfo{author}{\bibfnamefont{C.}~\bibnamefont{Kaur}}, \bibnamefont{and}
  \bibinfo{author}{\bibfnamefont{J.~D.} \bibnamefont{Weeks}},
  \bibinfo{journal}{J. Phys. Chem. B} \textbf{\bibinfo{volume}{108}},
  \bibinfo{pages}{19874} (\bibinfo{year}{2004}).

\bibitem[{\citenamefont{Denesyuk and
  Weeks}(2008)}]{DenesyukWeeks.2008.A-new-approach-for-efficient-simulation-of-Coulomb-interactions}
\bibinfo{author}{\bibfnamefont{N.~A.} \bibnamefont{Denesyuk}} \bibnamefont{and}
  \bibinfo{author}{\bibfnamefont{J.~D.} \bibnamefont{Weeks}},
  \bibinfo{journal}{J. Chem. Phys.} \textbf{\bibinfo{volume}{128}},
  \bibinfo{pages}{124109} (\bibinfo{year}{2008}).

\bibitem[{\citenamefont{Hansen and
  McDonald}(2006)}]{HansenMcDonald.2006.Theory-of-Simple-Liquids}
\bibinfo{author}{\bibfnamefont{J.-P.} \bibnamefont{Hansen}} \bibnamefont{and}
  \bibinfo{author}{\bibfnamefont{I.~R.} \bibnamefont{McDonald}},
  \emph{\bibinfo{title}{Theory of Simple Liquids}}
  (\bibinfo{publisher}{Academic Press}, \bibinfo{address}{New York},
  \bibinfo{year}{2006}), \bibinfo{edition}{3rd} ed.

\bibitem[{\citenamefont{Berendsen et~al.}(1987)\citenamefont{Berendsen,
  Grigera, and
  Straatsma}}]{BerendsenGrigeraStraatsma.1987.The-missing-term-in-effective-pair-potentials}
\bibinfo{author}{\bibfnamefont{H.}~\bibnamefont{Berendsen}},
  \bibinfo{author}{\bibfnamefont{J.}~\bibnamefont{Grigera}}, \bibnamefont{and}
  \bibinfo{author}{\bibfnamefont{T.}~\bibnamefont{Straatsma}},
  \bibinfo{journal}{J. Phys. Chem.} \textbf{\bibinfo{volume}{91}},
  \bibinfo{pages}{6269} (\bibinfo{year}{1987}).

\bibitem[{\citenamefont{Hu et~al.}(in preparation)\citenamefont{Hu, Rodgers,
  and
  Weeks}}]{HuRodgersWeeks.2008.On-the-efficient-short-ranged-simulations-of-polar-molecular}
\bibinfo{author}{\bibfnamefont{Z.}~\bibnamefont{Hu}},
  \bibinfo{author}{\bibfnamefont{J.~M.} \bibnamefont{Rodgers}},
  \bibnamefont{and} \bibinfo{author}{\bibfnamefont{J.~D.} \bibnamefont{Weeks}}
  (\bibinfo{year}{in preparation}).

\bibitem[{\citenamefont{Chandler}(1977)}]{Chandler.1977.Dielectric-constant-and-related-equilibrium-properties-of-molecular}
\bibinfo{author}{\bibfnamefont{D.}~\bibnamefont{Chandler}},
  \bibinfo{journal}{J. Chem. Phys.} \textbf{\bibinfo{volume}{67}},
  \bibinfo{pages}{1113} (\bibinfo{year}{1977}).

\bibitem[{\citenamefont{Stillinger and
  Lovett}(1968)}]{StillingerLovett.1968.General-Restriction-on-Distribution-of-Ions-in-Electrolytes}
\bibinfo{author}{\bibfnamefont{F.~H.} \bibnamefont{Stillinger}}
  \bibnamefont{and} \bibinfo{author}{\bibfnamefont{R.}~\bibnamefont{Lovett}},
  \bibinfo{journal}{J. Chem. Phys.} \textbf{\bibinfo{volume}{49}},
  \bibinfo{pages}{1991} (\bibinfo{year}{1968}).
 
\bibitem{Linse} D.~M. Ceperley and C.~V. Chester, Phys. Rev. A  \textbf{15},  755 (1977); P. Linse and H.~C. Andersen, J. Chem. Phys. \textbf{85}, 3027 (1986).

\bibitem[{\citenamefont{Smith}(2006)}]{Smith.2006.DLPOLY-applications-to-molecular-simulation-II}
\bibinfo{author}{\bibfnamefont{W.}~\bibnamefont{Smith}}, \bibinfo{journal}{Mol.
  Sim.} \textbf{\bibinfo{volume}{32}}, \bibinfo{pages}{933}
  (\bibinfo{year}{2006}).

\bibitem[{\citenamefont{Berendsen et~al.}(1984)\citenamefont{Berendsen, Postma,
  van Gunsteren, Dinola, and
  Haak}}]{BerendsenPostmaGunsteren.1984.Molecular-dynamics-with-coupling-to-an-external-bath}
\bibinfo{author}{\bibfnamefont{H.~J.~C.} \bibnamefont{Berendsen}},
  \bibinfo{author}{\bibfnamefont{J.~P.~M.} \bibnamefont{Postma}},
  \bibinfo{author}{\bibfnamefont{W.~F.} \bibnamefont{van Gunsteren}},
  \bibinfo{author}{\bibfnamefont{A.}~\bibnamefont{Dinola}}, \bibnamefont{and}
  \bibinfo{author}{\bibfnamefont{J.~R.} \bibnamefont{Haak}},
  \bibinfo{journal}{J. Chem. Phys.} \textbf{\bibinfo{volume}{81}},
  \bibinfo{pages}{3684} (\bibinfo{year}{1984}).

\bibitem[{\citenamefont{Chandler and
  Pratt}(1976)}]{ChandlerPratt.1976.Statistical-Mechanics-of-Chemical-Equilibria-and-Intramolecular}
\bibinfo{author}{\bibfnamefont{D.}~\bibnamefont{Chandler}} \bibnamefont{and}
  \bibinfo{author}{\bibfnamefont{L.~R.} \bibnamefont{Pratt}},
  \bibinfo{journal}{J. Chem. Phys.} \textbf{\bibinfo{volume}{65}},
  \bibinfo{pages}{2925} (\bibinfo{year}{1976}).

\bibitem[{\citenamefont{Martin}(1988)}]{Martin.1988.Sum-rules-in-charged-fluids}
\bibinfo{author}{\bibfnamefont{P.}~\bibnamefont{Martin}},
  \bibinfo{journal}{Rev. Mod. Phys.} \textbf{\bibinfo{volume}{60}},
  \bibinfo{pages}{1075} (\bibinfo{year}{1988}).

\bibitem[{\citenamefont{Hummer et~al.}(1996)\citenamefont{Hummer, Pratt, and
  Garcia}}]{HummerPrattGarcia.1996.Free-energy-of-ionic-hydration}
\bibinfo{author}{\bibfnamefont{G.}~\bibnamefont{Hummer}},
  \bibinfo{author}{\bibfnamefont{L.}~\bibnamefont{Pratt}}, \bibnamefont{and}
  \bibinfo{author}{\bibfnamefont{A.}~\bibnamefont{Garcia}},
  \bibinfo{journal}{J. Phys. Chem.} \textbf{\bibinfo{volume}{100}},
  \bibinfo{pages}{1206} (\bibinfo{year}{1996}).
  
\bibitem{pressurecorrect} R. Topol  and P. Claverie,  Mol. Phys. \textbf{35}, 1753 (1978);
D. Chandler in \emph{The Liquid State of Matter: Fluids, Simple and Complex}
(North Holland, New York, 1982), p. 275.

\bibitem[{\citenamefont{Percus}(1964)}]{Percus.J.The-pair-distribution-function-in-classical-statistical.1964}
\bibinfo{author}{\bibfnamefont{J.~K.} \bibnamefont{Percus}}, in
  \emph{\bibinfo{booktitle}{The Equilibrium Theory of Classical Fluids}}
  (\bibinfo{publisher}{W.~A.~Benjamin, Inc.}, \bibinfo{address}{New York},
  \bibinfo{year}{1964}), p. II-33.

\end{thebibliography}

\end{document}